 \documentclass[final,5p,times,twocolumn]{elsarticle}

 \usepackage{graphics}
\usepackage{url}

\newcommand{\CL}{\mathcal{L}}
\newcommand{\CP}{\mathcal{P}}

\usepackage{amssymb}

\biboptions{comma,square,sort&compress}

\journal{Computer Networks}

\begin{document}

\begin{frontmatter}



\title{Large scale probabilistic available bandwidth estimation}


\author[mcgill]{Frederic Thouin\corref{cor1}}
\ead{frederic.thouin@mail.mcgill.ca}

\author[mcgill]{Mark Coates}
\ead{mark.coates@mcgill.ca}

\author[mcgill]{Michael Rabbat}
\ead{michael.rabbat@mcgill.ca}

\cortext[cor1]{Corresponding author. Tel.: +1-514-398-5516.  Fax: +1-514-398-3127}

\address[mcgill]{McGill University, Department of Electrical and Computer Engineering, 3480 University, Montreal, Quebec, Canada, H2A 3A7}

\begin{abstract}

  The common utilization-based definition of available bandwidth and many of the existing tools to estimate it suffer from several important weaknesses: i) most tools report a point estimate of average available bandwidth over a measurement interval and do not provide a confidence interval; ii) the commonly adopted models used to relate the available bandwidth metric to the measured data are invalid in almost all practical scenarios; iii) existing tools do not scale well and are not suited to the task of multi-path estimation in large-scale networks; iv) almost all tools use ad-hoc techniques to address measurement noise; and v) tools do not provide enough flexibility in terms of accuracy, overhead, latency and reliability to adapt to the requirements of various applications.  In this paper we propose a new definition for available bandwidth and a novel framework that addresses these issues.  We define {\em probabilistic available bandwidth} (PAB) as the largest input rate at which we can send a traffic flow along a path while achieving, with specified probability, an output rate that is almost as large as the input rate.  PAB is expressed directly in terms of the measurable output rate and includes adjustable parameters that allow the user to adapt to different application requirements.  Our probabilistic framework to estimate network-wide probabilistic available bandwidth is based on packet trains, Bayesian inference, factor graphs and active sampling.  We deploy our tool on the PlanetLab network and our results show that we can obtain accurate estimates with a much smaller measurement overhead compared to existing approaches.  \end{abstract}


\end{frontmatter}

\section{Introduction}
\label{sec:intro}

Recent work has shown that the performance of applications such as overlay network routing~\cite{hir:07,lee:08} and anomaly detection~\cite{he:08anomaly} can be improved significantly when the network-wide available bandwidth is known.  There are many more applications (SLA compliance, network management, transport protocols, traffic engineering, admission control) that could also benefit from this information, but existing tools that measure available bandwidth generally do not meet the requirements of these applications in terms of accuracy, overhead, timeliness and reliability~\cite{gue:09applicability}.  

The most popular estimation tools are founded on either the probe-gap (PGM) or probe-rate model (PRM).  The PGM assumes a single-hop path with FIFO queuing and fluid cross-traffic\footnote{Traffic is modelled as a continuum of infinitely small packets with an average rate that changes slowly.}.  One measurement consists of sampling cross-traffic by observing the gap between a packet pair at both the input and the output.  With every measurement, a single point estimate of the available bandwidth can be produced as long as i) the capacity of the tight link is known, ii) there is only one tight link and it is the same as the narrow link and iii) the end-nodes can transmit faster than the available bandwidth.  PGM-based tools (e.g., Spruce~\cite{str:03}, IGI~\cite{hu:03}) are lightweight and fast, but are unable to estimate the available bandwidth of multi-hop paths~\cite{lao:06}.  
The probe-rate model (PRM) also assumes fluid cross-traffic, but is more robust.
The PRM relies on the principle of self-induced congestion probing~\cite{rib:03}: if probes are sent at a rate smaller than the available bandwidth then the output rate matches the probing rate.  However, if the probing rate is greater than the available bandwidth, packets get queued, which results in unusual delays and a smaller output rate.  Algorithms constructed using the PRM (e.g., Pathload~\cite{jai:03}, pathChirp~\cite{rib:03}) consist of varying the probing rate to identify the boundary that separates the two different behaviours described above: an input rate where probes start experiencing unusual delays.  These methods generate more accurate estimates than PGM-based tools, but they are also more intrusive because they require multiple iterations at different probing rates.

In addition to the lack of flexibility, existing models and tools suffer from four other major weaknesses:
\begin{enumerate}
	\item The vast majority report a single value representing average available bandwidth and the usefulness of this single value is questionable. Available bandwidth is typically defined as the capacity of a path unused by cross-traffic over a specified time period. 
Most tools produce a single point estimate of the available bandwidth by making multiple measurements using probes sent throughout the time period of interest.  The cross-traffic often fluctuates significantly over the time period, so probes experience very different network conditions; an estimate formed from such data can be a high-variance quantity making a confidence interval very valuable.
	Service (or response) curves are more informative than single average estimates; they present the statistical mean (asymptotic average) of the output rate for an entire range of input rates~\cite{liu:08}. However, each point on the curve is still an average that does not really provide a meaningful reflection of the burstiness of the traffic and the variability of the available bandwidth metric. A more robust and practically-relevant manner to express the available bandwidth is the variation range (confidence interval) proposed by \citet{jai:03}.  
	\item The observation model relating measured data to the utilization-based definition of available bandwidth is inaccurate and biased in most practical situations. As a result, the value provided by most tools does not genuinely reflect the quantity the tools claim to estimate.  The fluid cross-traffic assumption underpins the vast majority of models used for inference. \citet{liu:08} show that the assumed relationships between the measured quantities (packet dispersion, one-way delay, output rate) and the estimated value (utilization, unused capacity) are not sound; even for simple, slightly more realistic scenarios, the adoption of a fluid model leads to significant underestimates of the available bandwidth (unused capacity).
	\item The mechanisms used by most tools to handle measurement noise are ad-hoc and, in many cases, inadequate.
Measurement errors and noise generated by the end-hosts and routers along the end-to-end path are unavoidable in practice. Common issues include route changes, out-of-order packet delivery, packet replications, errors in the probing packets due to link quality issues, incorrect packet time stamps, and poor Network Interface Card utilizations. Although measures can be adopted to prevent some of these errors, it is impossible to eradicate them all. It is important that the model and inference technique are robust, and that they can tolerate and handle noisy measurements. One example of a technique that does handle noise more robustly is Traceband~\cite{gue:09traceband}, which employs a hidden Markov model that allows the technique to statistically adjust to noise in the measurements.
	\item Current tools cannot be applied to larger networks to simultaneously estimate the available bandwidths of multiple paths.  Using existing tools, probing all paths concurrently not only introduces an unacceptable overhead and overloads hosts, but also leads to significant underestimation due to interference between the probes on links shared by multiple paths~\cite{cro:09}. The alternative to simultaneous measurements is to sequentially probe each path independently. This is unacceptably time-consuming and very inefficient, however, because it ignores the significant correlations that arise in available bandwidth metrics when the network paths share links.
\end{enumerate}

In this paper, we tackle the problem of network-wide (multi-path) available bandwidth estimation. In developing our approach, we strive to address the issues we have identified above. This problem can be related to large-scale network inference. There are similarities with network tomography\footnote{See~\cite{cas:04} and references therein for a review of network tomography.}, which consists of estimating either i) link-level parameters based on end-to-end measurements; or ii) path-level traffic intensity based on link-level traffic measurements~\cite{var:96}.  There are two key differences. First, tomography involves a mapping from path-level measurements to link-level metrics or vice versa; in the network-wide available bandwidth problem we are interested in estimating path-level metrics from path-level measurements. Second, in most network tomography problems, there is a linear relation of the form $y=Ax$ between measurements $y$ and network parameters $x$, where $A$ is a routing matrix. In our problem, 
this relationship is non-linear; one of our modelling assumptions is that the available bandwidth of a path is the minimum of the available bandwidths of all its constituent links. 

The task is more closely related to the problem of {\em network kriging}~\cite{chu:06}, which involves estimating (functions of) path-level metrics throughout a network using end-to-end path measurements. This problem was also addressed in~\cite{che:03,che:07algebra}, where an algebraic approach was proposed for exactly recovering, under the assumption of no noise, the path level metrics of all the end-to-end paths in a network by monitoring only a small subset of the paths. The method in~\cite{chu:06} reduced this monitoring cost even further, at the expense of introducing a small error in the estimated metrics.  
For real-time applications, estimates must not only be produced with minimal overhead, but also in a timely manner.  To meet these requirements, measurements, even for a reduced subset of paths, must be scheduled at the same time.  
To avoid simultaneous probes interfering with each other and overloading nodes,~\citet{son:07} propose a resource-aware technique that achieves better accuracy than resource-oblivious methods at the cost of using more measurement data.
All of these approaches, as well as the wavelet-based methodology described in~\cite{coa:07}, are only appropriate for (approximately) additive metrics, such as loss or delay, where a linear relationship can be constructed between the link-level and path-level metrics.
However,~\citet{son:07} suggest that their approach could be extended to available bandwidth estimation by selecting paths such that the load of their probes only represents a small fraction of the capacity of each link.


Large-scale (multi-path) estimation of available bandwidth has not received as much attention as other metrics.
To limit measurement overhead, BRoute~\cite{hu:05} capitalizes on the spatial correlation between links shared by many paths and the observation that $86\%$ of Internet bottleneck links are within four hops (end-segments) from end nodes~\cite{hu:04}.  The tool first uses traceroute landmarks to identify AS-level end segments for each node, and then measures available bandwidth on these segments
by using landmarks with high downstream bandwidth.  \citet{man:07bandwidth} propose a more efficient landmark-based approach that is similar to BRoute but has reduced storage and inference complexity.
Another approach to large scale available bandwidth estimation is to exploit the correlation between various metrics (route, number of hops, capacity and available bandwidth); since the measurement cost for each metric is different, monitoring those that have a cheaper cost can reduce the load on the network~\cite{yal:08}.  
To further reduce the amount of probing overhead,~\citet{man:07inferring} propose to reshape existing TCP traffic to look like packet pairs, trains or chirps so that no extra traffic is injected in the network.  
Despite these efforts to minimize the overhead of the estimation procedure, most of these network-wide tools do not address any of the concerns mentioned earlier; they are neither flexible nor robust to noisy measurements, they produce a single average value for each path and they are based on an invalid mapping between measurements and the inferred metrics.

\subsection{Contributions}

We present a novel system that addresses the five weaknesses discussed above.  Our solution includes i) a probabilistic-rate-base definition for the available bandwidth and ii) a network-wide estimation tool.  

Our implementation uses the Bayesian inference framework, factor graphs and the belief propagation algorithm to fuse the information obtained from all measurements.
We adopt a model that relates the PAB of each path to the PAB of its constituent links; the factor graph provides a mechanism for capturing this model and enables computationally efficient inference.  These techniques have been successfully used in large-scale network problems, such as link loss inference applications~\cite{coa:00,mao:05} and the computation of conditional entropies for both fault diagnosis and most informative test selection~\cite{ris:05distributed,zhe:05,ris:06}, but not yet in the context of available bandwidth estimation.  

Another novel contribution is our algorithm to determine which path and rate to probe at each iteration; 
a process that can be related to sequential Bayesian sampling~\cite{elg:93} and active/adaptive sampling~\cite{cas:07}.  
This sampling strategy consists of selecting the next measurement(s) based on the information acquired previously, such that the expected information gain is maximized.  
In networking, it has been used in the context of network tomography to determine the measurements that provide the best information gain about the network path property given their probing overhead~\cite{son:09}, but has yet to be applied to available bandwidth estimation.

The rest of this paper is organized as follows.  In Sect.~\ref{sec:pab}, we introduce a new metric, {\em probabilistic available bandwidth}, and formally state the estimation problem.  In Sect.~\ref{sec:meth}, we detail our novel probabilistic framework, which is the first to combine factor graphs and active sampling to estimate available bandwidth.  In Sect.~\ref{sec:res}, we present results from our simulations and online experiments on the PlanetLab network.  In Sect.~\ref{sec:conclusion}, we summarize our contributions and discuss future work.

\section{Probabilistic Available Bandwidth}
\label{sec:pab}

%
%
%

We specify the {\em probabilistic available bandwidth} (PAB) metric directly in terms of input rates and output rates of traffic on a path.  We are interested in determining the largest input rate $r_p$ at which we can send a traffic flow along a path while achieving an output rate $r'_p$ that is almost (within $\epsilon$) as large as the input rate, with specified probability\footnote{The probability is defined over all possible multi-packet flows of average rate equal to the input rate that can complete transmission during the specified measurement period.} at least $\gamma$.  More formally, for given $\epsilon > 0$ and $\gamma > 0$, we seek the largest input rate such that $\Pr(r'_p > r_p - \epsilon) \geq \gamma$.  We denote the largest such ingress rate by $y_p$ and refer to it as the probabilistic available bandwidth for path $p$: $$y_p = \max_{r_p} \Pr(r'_p > r_p - \epsilon) \geq \gamma.$$  The probabilistic available bandwidth is located at the boundary of two regions with different behaviours (i.e., where we can expect different outputs).  For smaller rates,  $r_p \leq y_p$, there is a probability greater or equal to $\gamma$ that the output rate will be within a margin of $\epsilon$ of the input rate.  For input rates greater than the PAB, $r_p>y_p$, this probability is not guaranteed.  

We believe that this new definition for available bandwidth is more robust and practical for several important reasons.  First, it provides a more valid mapping between the measured and inferred quantities.  By expressing available bandwidth directly in terms of the input and output rates, there is no longer a need to bridge the gap between packet dispersion and unused capacity through generally invalid modelling assumptions.  Second, the probabilistic framework gives flexibility to the user and is more resistant to variability (cross-traffic burstiness) and noise (errors) in the measurements.  The values of the two parameters $\epsilon$ and $\gamma$ are defined by the user based on application requirements and the network environment.  For example, increasing the value of $\gamma$ results in a more conservative (smaller) estimate of the probabilistic available bandwidth.  In a network where frequent measurement errors occur, the value of $\epsilon$ can be increased, if the application can tolerate a certain reduction in output rate.  Last, it represents a more practical and concrete quantity: the probability that transmitting data at a given rate will yield the desired (same) output rate.

\subsection{Problem Statement}
\label{ssec:ps}

We focus on the problem of network-wide available bandwidth estimation, but in terms of our newly introduced metric, probabilistic available bandwidth.  More formally, for a specified $(\epsilon,\gamma)$ and network that consists of a set of $N$ links $\CL$ and $M$ paths $\CP$, we wish to form estimates of the probabilistic available bandwidths of all paths in the network.  Let the PAB of each path $p$ be modelled as a discrete\footnote{We chose to define $y_p$ as a discrete, rather than continuous, random variable because it  not meaningful to have an infinite precision on the transmission rates.} random variable $y_p$; e.g., $\Pr(y_p = r)$ being the probability that the PAB on path $p$ is $r$.  

We use an iterative probing strategy where, for each measurement, we wish to determine if the probing rate is greater or smaller than the probabilistic available bandwidth.  At each iteration $k$, we evaluate a binary outcome\footnote{Despite the loss of information, we choose to produce a binary outcome rather than use the output rate directly for two reasons.  First, a binary outcome is more robust and less sensitive to noisy measurements.  Second, there is no available likelihood model for the output rate and it is easier to construct empirically an accurate one for the binary outcome.} $z_k$ that specifies whether the egress rate was within $\epsilon$ of the ingress rate.  Then at any given instant $k$, we are interested in the marginal posterior $\Pr(y_p | \mathbf{z})$ for every path $p$, where $\mathbf{z}= [z_1,\dots, z_k]$.  Our goal is to identify a probing method and the most informative measurement at each iteration in order to form the PAB estimates, such that the credible intervals of the estimates (based on the marginal posteriors) are acceptably tight and the measurement overhead is minimal.

\begin{figure}[!h]
	\centering
	\includegraphics[width=\linewidth]{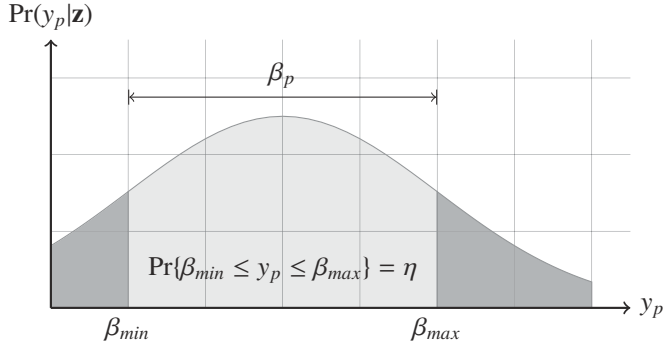}
	\caption{Graphic representation of the probabilistic available bandwidth.  The probability that $y_p$ lies in the confidence range $[\beta_{min},\beta_{max}]$ of size $\beta_p$ is equal to $\eta$ (confidence level).
\label{fig:confidence_interval}}
\end{figure}

Rather than estimating the PAB by a single value, we identify a confidence interval likely to include it.  For a given distribution, such as the one depicted in Fig.~\ref{fig:confidence_interval}, the confidence interval of size $\beta_p$ with confidence limits $[\beta_{min},\beta_{max}]$ is the smallest interval that has a confidence level (fraction of probability mass) greater than $\eta$.  The estimation procedure terminates when the size of the confidence interval of each path is smaller than $\beta$ ($\forall p: \beta_p \leq \beta$).  For the cases when the variability of the measurements is too high to meet the desired tightness for confidence intervals, the procedure also stops when the maximum number of iterations is reached.

The value of $\eta$ and the desired size of the confidence interval $\beta$ (how tight the interval is) are both defined by the user depending on application requirements and determine how accurate, fast and intrusive the estimation tool is.  For example, a larger $\beta$ or smaller $\eta$ will generally require a smaller number of measurements, which leads to a faster estimation with a smaller overhead, but also a less accurate one.  It is important to understand the distinction between $\gamma$ and $\eta$.  The confidence level for a path $\eta$ is the probability that $y_p$ lies in the confidence interval of size $\beta_p$ bounded between $\beta_{min}$ and $\beta_{max}$.  The probability of success $\gamma$ represents the probability, for rates smaller than the probabilistic available bandwidth $y_p$, that the output rate is within a margin of $\epsilon$ of the input rate.

\section{Methodology}
\label{sec:meth}

\begin{figure}[!h]
	\centering
	\includegraphics[width=\linewidth]{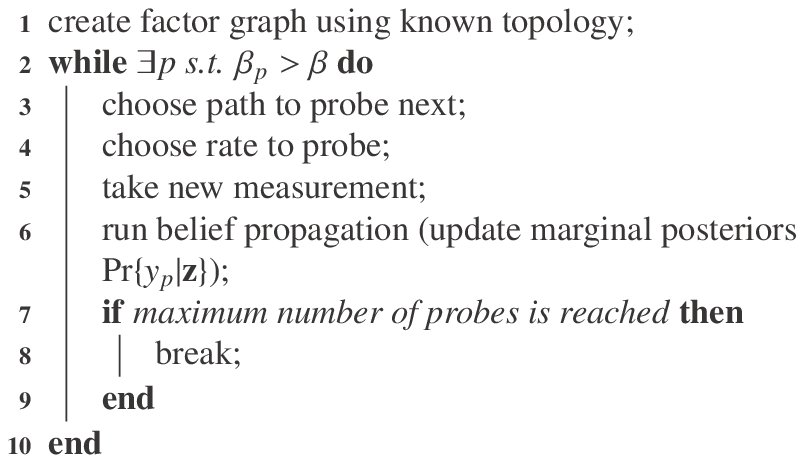}
	\caption{Multipath probabilistic available bandwidth estimation algorithm.\label{fig:alg_main}}
\end{figure}

Our main challenge is to develop a technique to estimate probabilistic available bandwidth that is efficient and scales well with the number of paths.  
We can divide this problem into the following three tasks: i) measure a path and produce a binary outcome, ii) compute the marginal of the path's probabilistic available bandwidth from measurement outcomes and establish confidence intervals for the PAB, and iii) identify  measurements (choose the path and probing rate) at each iteration that will minimize the overhead on the network.  A general overview of our approach is presented in Fig.~\ref{fig:alg_main}.  We will explain each line (except for the termination criteria of line 2 and 7 presented in Sect.~\ref{ssec:ps}) in the rest of this section. 

Our method is based on four assumptions.  
\begin{enumerate}
\item At the start of each link is a store-and-forward first-come first-served router/switch that dictates the behaviour of the link (in terms of delay, loss, utilization).  If the network uses priority queueing or some other form of router-level Quality-of-Service provisioning, then our method will infer the probabilistic available bandwidth as seen by the class of packets transmitted as probes.
\item The routing topology of this network is known, as embodied in the set of paths $\CP$, and that it remains fixed for the duration of our experiments.  
More precisely, we construct a $MxN$ binary path matrix $\mathbf{P}$, where $\mathbf{P}(i,j)$ is equal to one if link $j$ is on path $i$.  
To populate the matrix, we infer links and the mapping from IP addresses to routers using $\texttt{traceroute}$\footnote{\texttt{traceroute}-like methods have been known to inflate the number of observed routers, record incorrect links and bias router degree distributions~\cite{she:08}.  However, it provides sufficiently accurate topology estimates for us to assess the performance of our algorithms.}.  
\item There is a unique path between each of the hosts involved in probing.  If there is per-packet load balancing in the network, our $\texttt{traceroute}$-based procedure will identify only one of these paths traversed by packets.  This error takes the form of missing correlations in the factor graph and could result in inaccurate estimates and/or slower convergence.  Our method is unaffected by destination-based load balancing.
\item Like the majority of utilization-based available bandwidth estimation tools, we assume that there is a single link (tight link) on each path that essentially determines the probabilistic available bandwidth of that path.  More formally, each path consists of the set of links $L_p = \{\ell_1,\ell_2,\dots,\ell_n\}$ and a single tight link ${\ell}^* \in L_p$\footnote{We derive this relationship more formally in Sec.~\ref{sssec:path_link_relation}.}.  This allows us to i) perform efficient inference using path-level data and ii) use logical topologies (combine all links that are in a series) rather than routing topologies to reduce the number of links and the complexity of the factor graph.~\citet{jai:03} show that multiple tight links can lead to an underestimation of the available bandwidth.  In our case, we interpret the presence of more than one tight link as a modelling inaccuracy that creates noise propagated in the factor graph during the execution of the belief propagation algorithm.
\end{enumerate}
	
We revisit these assumptions in Sec.~\ref{ssec:robust} and study how errors or changes in routing topology affect the performance of our algorithm.			
	
\subsection{Probing Strategy}

Our probing strategy (line 5 in Fig.~\ref{fig:alg_main}) is a based on the principle of self-induced congestion~\cite{rib:03}.  A single measurement consists of sending $N_t$ trains of $L_s$ UDP packets of $P_{size}$ bytes at a constant rate $r_p$ and observing the rate $r'_p$ at the receiver side.  We then take the median of $r'_p$ obtained from each of the $N_t$ trains and determine the binary outcome $z$ of the measurement using the following relation: $z = \mathbf{1}\{r'_p \geq r_p-\epsilon\}$ where $\mathbf{1}(x)$ is the indicator function (equal to one if $x$ is true and zero otherwise).  

To achieve a given input rate $r_p$, we fix the packet size and calculate the time interval, $\tau$, between the departure of consecutive packets according the the following relation: $r_p = P_{size}/\tau$.  The receiving rate is calculated similarly by dividing the total number of bytes received by the amount of time that elapsed between the reception of the first and last packet. However, due to task interruption on the sender side there can be unusual delays between the departure of two consecutive packets ($t_i > t_{i-1} + \tau$ where $t_i$ is the departure time of packet $i$).  We consider these packets invalid and exclude them before calculating the output rate.  Upon reception of the last packet of a train, we construct a set $V$ of all the indices $i>1$ of valid packets and calculate $r'_p$ as follows: $r'_p = (|V| \cdot P_{size})/\left(\sum_{i \in V} t_i - t_{i-1}\right)$.

The probing rate is selected at every iteration, but the other parameters are pre-determined before the beginning of the estimation procedure.  The choice of these values is made to minimize the overhead while making sure that results are accurate.  In active sampling techniques, the outcome of each measurement plays a role in determining what rate to probe next.  Although using multiple trains ($N_t > 1$) and taking the median of the output rates increases the overhead on the network, it is also a way to mitigate the impact of a  noisy measurement sequence (e.g. packet train with many invalid packets).  A similar logic applies when choosing the size of each probe, $P_{size}$, and the number of probes in a train, $L_s$.  Larger probes and longer trains provide more samples over which to average $r'_p$, but also leads to a more significant load on the network and a longer sampling period.  In the Sec.~\ref{sec:res}, we specify and justify our choices for each of these parameters.

\subsection{Bayesian Inference and Factor Graphs}
\label{ssec:bay_inf_fac_gra}

Bayesian inference is a classical way to update the knowledge about unknown parameters based on new observations.  
In this framework, the posterior distribution $\Pr(y_p | z_k)$ is proportional to the product of the conditional probability $\Pr(z_k | y_p)$, also called likelihood function, and the prior probability $\Pr(y_p)$: $\Pr(y_p | z_k) \propto \Pr(z_k | y_p) \Pr (y_p)$.  We are interested in the marginal, for every path, of the joint posterior distribution of all paths $\Pr(y_1,...,y_M | \mathbf{z})$.  
The joint probability distribution is complex but it is factorizable and can therefore be captured with a factor graph (line 1 in Fig.~\ref{fig:alg_main}): a graphical model ``that indicates how a joint function of many variables factors into a product of functions of smaller sets of variables"~\cite{fre:98}.  
Factor graphs are composed of two types of nodes (variable and factor nodes) and edges that show dependencies between the variables and the factors.
In our case, the variables are discrete random variables of the probabilistic available bandwidth of each link, $x_{\ell}$, and path, $y_p$.  There are three functions that are represented by factor nodes in the graph: i) the prior knowledge about the links, $f_x$, ii) the relation between the PAB of links and paths, $f_{x,y}$; and iii) the likelihood of an observation on a given path, $f_{y,z}$.

The marginal posteriors are computed (line 6 in Fig.~\ref{fig:alg_main}) by running belief propagation on the factor graph~\cite{pea:88}. The algorithm starts with each one of the leaf nodes (prior and likelihood) sending a message to its adjacent node.  Messages are then computed using the sum-product algorithm and continue to propagate until the algorithm stabilizes, i.e. there is minimal or no variation between a newly computed message and the one previously sent of the same edge\footnote{Belief propagation will converge in cyclic factor graphs under certain conditions, but is not guaranteed to do so~\cite{moo:07}.  Through our extensive simulations, we did not encounter any convergence issues.  To ensure completion, we set the maximum number of messages between two nodes to five during one run of the belief propagation algorithm.}.  Upon completion it is possible to compute the marginal at the variable node (links and paths) by taking the product of all messages incoming on its edges.

\begin{figure}[!h]
\centering
	\includegraphics[width=0.5\linewidth]{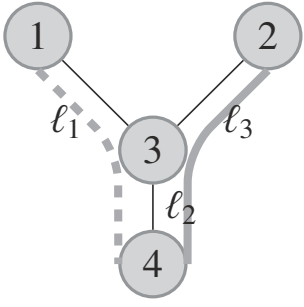}
	\caption{Logical topology of a 4 nodes network with $N=3$ links ($\ell_1,\ell_2,\ell_3$) and $M=2$ paths; $p_1$ (dashed) and $p_2$ (solid).  \label{fig:logicaltopology}}
\end{figure}

\begin{figure}
\centering
			\includegraphics[width=0.6\linewidth]{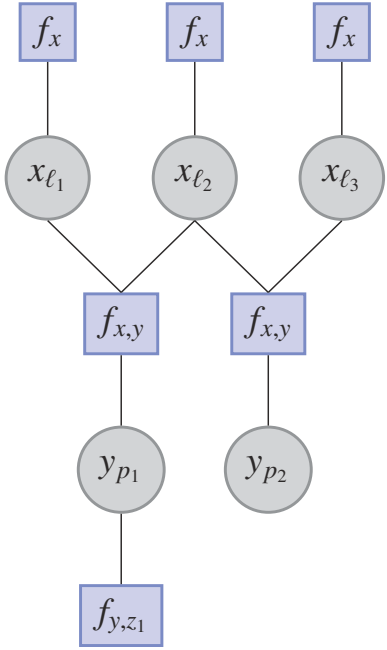}
\caption{Factor graph representation used to estimate the PAB of the two paths in the topology depicted in Fig.~\ref{fig:logicaltopology}.\label{fig:factor_graph}}
\end{figure}

{\bf Example:} In Figure~\ref{fig:logicaltopology}, we show an example of a simple logical topology of a network.  In this example, there are four nodes interconnected using $N=3$ different links labeled $\ell_1,\ell_2,\ell_3$ and we consider $M=2$ paths (dashed line: $p_1$, solid line: $p_2$) where nodes 1 and 2 are the sources and node 4 is the destination.  From the logical topology, we can populate the path matrix $\mathbf{P}$ and use it to construct the factor graph.
$$\mathbf{P} = \left[\begin{array} {ccc} 1 & 1 & 0 \\ 0 & 1 & 1\end{array} \right]$$
In Figure~\ref{fig:factor_graph}, we show the factor graph representation of the joint distribution used to compute marginal posteriors of the PAB of each of the three links and two paths.  The edges show the variables that the factors depend on.  In this case, the prior function is identical for all links.  So each variable node $x_\ell$ is connected to a factor node $f_x$ in the graph.  However, we could easily use different functions for each link.  Each path and its underlying set of links $L_p$ are connected together to a factor node $f_{x,y}$ (there is an edge for every $\mathbf{P}(i,j)=1$ in the path matrix).  Finally, we see that this specific factor graph includes information from a single observation that was performed on path $p_1$.  For each additional measurement, a new factor node $f_{y,z_k}$ is added to the factor graph.

\subsubsection{Prior function}
The first function to define is the prior $f_x$.  We use a non-informative prior model for the PAB of a path; a uniform distribution in the range $[B_{min},B_{max}]$: $$f_x \sim \mathcal{U}[B_{min},B_{max}],$$where $B_{\min}$ and $B_{\max}$ are conservative estimates of the minimum and maximum probabilistic available bandwidths of links.  Our choice is due to the lack of any prior information about the PAB of links or paths.

\subsubsection{Relation between links and paths}
\label{sssec:path_link_relation}

Our inference procedure relies on a relationship between the PAB of a path and the PABs of its constituent links.  For the classical utilization-based definition of available bandwidth, it is often assumed that there is a single link on each path that determines that path's available bandwidth.  We develop a similar relationship for the probabilistic available bandwidth.


For a path $p$ consisting of the set of links $L_p = \{1,2,\dots,n\}$, it is possible to identify small constants $0<\epsilon_{\ell} < \textstyle{\sum}_{{\ell} \in L_p} \epsilon_{\ell} < \epsilon$ and $0<\delta_{\ell} < \textstyle{\sum}_{{\ell} \in L_p} \delta_{\ell} < 1 -\gamma$ such that:

\begin{equation}
\Pr(r'_{\ell} \leq r_{\ell} - \epsilon_{\ell}) \leq \delta_{\ell} \quad {\mathrm{ for\,\, all }} \quad r_\ell \leq y_p(\epsilon,\gamma). \label{eq:linkcond}
\end{equation} 
but 
\begin{equation}
\Pr(r'_{\ell} \leq r_{\ell} - \epsilon_{\ell}) > \delta_{\ell} \quad {\mathrm{ for\,\, all }} \quad r_\ell > y_p(\epsilon,\gamma). \label{eq:linkcond2}
\end{equation} 

We can apply the union bound on the links to establish:
\begin{equation}
\Pr\left(\bigcup_{{\ell} \in L_p} \{r'_{\ell} \le r_{\ell} - \epsilon_{\ell}\}\right) \le \sum_{{\ell} \in L_p} \delta_{\ell}. \label{eq:unionbound}
\end{equation} 

The complement of this union bound is that the condition $r'_{\ell}>r_{\ell}-\epsilon_\ell$ holds for each link. Then we have the following relationship between the path and link input and output rates:
\begin{eqnarray*}
r_1 &=& r_p\\
r_2 &=& r_1' > r_p - \epsilon_1 \\
r_3 &=& r_2' > r_p - \epsilon_1 - \epsilon_2 \\
&\vdots& \\
r'_p &=& r'_n > r_p - \sum_{i=1}^n \epsilon_i.
\end{eqnarray*}

This relationship and the union bound in (\ref{eq:unionbound}) imply the following:
\begin{equation}
\Pr\left(r'_p > r_p - \sum_{{\ell} \in L_p} \epsilon_{\ell}\right) \geq 1 - \sum_{{\ell} \in L_p} \delta_{\ell}. \label{eq:pathcond}
\end{equation}


Moreover, we assume that there is a \emph{tight link} ${\ell}^* \in L_p$ which essentially determines the probabilistic available bandwidth on the path $p$. 
This means that it is possible, for all ${\ell}\in L_p$, ${\ell} \ne {\ell}^*$, to identify $\epsilon_{\ell}\ll \epsilon$ and $\delta_{\ell}\ll 1 - \gamma$ that satisfy (\ref{eq:linkcond}). 
In the case of ${\ell}^*$, however, the smallest $\epsilon_{{\ell}^*} < \epsilon$ and $\delta_{{\ell}^*} <1 - \gamma$ pair that satisfy (\ref{eq:linkcond}) have the property $\epsilon_{{\ell}^*} \approx \epsilon$ and $\delta_{{\ell}^*} \approx 1 - \gamma$.
The tight link assumption implies that $\textstyle{\sum}_{{\ell} \in L_p} \epsilon_{\ell} \approx \epsilon_{{\ell}^*} \approx \epsilon$ and $\textstyle{\sum}_{{\ell} \in L_p} \delta_{\ell} \approx \delta_{{\ell}^*} \approx 1 - \gamma$. 
This property, together with (\ref{eq:linkcond}), (\ref{eq:linkcond2}), and (\ref{eq:pathcond}), imply that $y_p \approx x_{{\ell}^*}$ where $x_{\ell}$ is the PAB of link $\ell$. 
Another way of interpreting this assumption, is that the PAB of any link $\ell \in L_p$, $\ell \neq {\ell}^*$ is significantly greater than $y_p$. 
This relationship is expressed mathematically as $$f_{x,y} (y_p, \{x_{\ell} | \ell \in L_p\}) = \mathbf{1} \{y_p = \min_{\ell \in L_p} (x_\ell)\},$$ where $\mathbf{1}\{x\}$ is the indicator function. 

\subsubsection{Likelihood Model}
\label{sssec:likelihood}

\begin{figure}[!h]
	\centering
	\includegraphics[angle=-90,width=\linewidth]{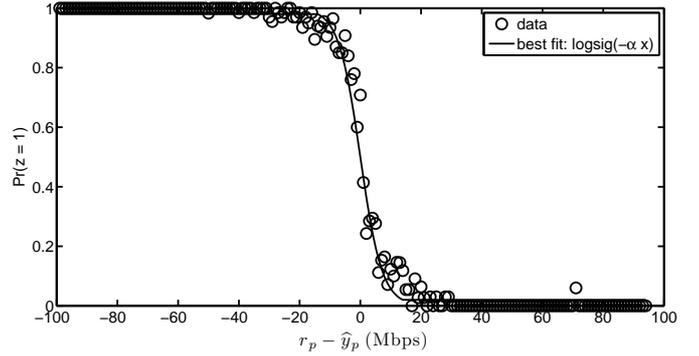}
	\caption{Empirical data and regression fit for the likelihood model.  $Pr(z = 1)$ is a function of the difference between the probing rate and estimated available bandwidth.  Each data point is obtained by averaging the result of 10 packet trains with $\epsilon=5$ over five different paths.  The best fit is obtained by performing a regression for parameters $\alpha$ and $y_p$.\label{fig:likelihood_info}}
\end{figure}

Each measurement $k$ is a $(p,r_{p}^k,z_{k})$ triple that consists of the outcome $z_{k}$, the probed path $p$ and probed rate $r_{p}^k$.  We specify a likelihood function, $f_{y,z}$, learned from empirical training data, that relates this outcome to the probe rate and the underlying PAB of the probed path.  This function depends on the probing strategy and how the outcome of a measurement is determined.  Intuitively, when the probing rate $r_p$ is well below $y_p$, we expect the probability of observing $z=1$ to be very high and, similarly, when $r_p$ is well over $y_p$, this probability should be very close to zero.  Although a simple step function looks like a good match, it is too aggressive as we have observed higher levels of noise when we probe around $y_p$.  Based on these intuitive expectations and experimental data (Fig.~\ref{fig:likelihood_info}), we adopt the likelihood model $$L(z=1|y_p,r_p)={\mathrm{logsig}}(-\alpha(r_p-y_p))$$ for the measurements, where $\alpha$ is a small positive constant learned empirically\footnote{The sigmoid function rapidly decays to zero when the probing rate is greater than the available bandwidth, even for the best possible parameter fit.  We wish to be careful and prefer a slightly less aggressive approach where we assign some likelihood to unexpected measurement outcomes at all ingress rates.  For that reason, we introduce a small constant $\kappa$ and bound our likelihood function to lie in the range $[\kappa ,1-\kappa ]$; in our experiments $\kappa = 0.02$.}.  However, to determine the value of $\alpha$ we first need to estimate $y_p$.  We decide to co-jointly estimate the values of $y_p$ along with the constant $\alpha$ through a single regression procedure where we determine the best fit by minimizing the MSE. 

We note that our estimation procedure is not sensitive to the exact choice of $\alpha$, which specifies the rate of decay of the sigmoid function. Moreover, in experiments conducted on different topologies, days, and times-of-day, we have observed that the estimated $\alpha$ values occupy a small range. The values are related to the variability of the path PABs over the measurement interval. These observations suggest that it is possible to execute the training procedure rarely.

{\bf Example:} We construct a likelihood model for the network we used for our experiments using $\epsilon = 5$ Mbps and a range of values where $B_{min} = 1$ Mbps and $B_{max} = 100$ Mbps.  We first gather data from five different paths: 500 measurements from non-consecutive packet trains at each rate between $B_{min}$ and $B_{max}$.  We then repeat this experiment five times at different periods of the day resulting in 25 sets of 500 measurements.  We normalize each of the 25 experiments and combine all the data in a single plot as a function of $r_p - \widehat{y}_p$. The result is shown in Fig.~\ref{fig:likelihood_info} where each data point is the result of averaging all values which had the same value of $r_p - \widehat{y_p}$; all experiments for which the distance between $r_p$ and $y_p$ is identical.  The function depicted is for $\gamma = 0.5$, but it can be easily modified for any other value of $\gamma$: it consists of aligning the desired value of $\gamma$ on the curve with the point on the x-axis where $r_p - \widehat{y}_p = 0$.

\begin{figure}[!h]
\centering
\includegraphics[width=\linewidth]{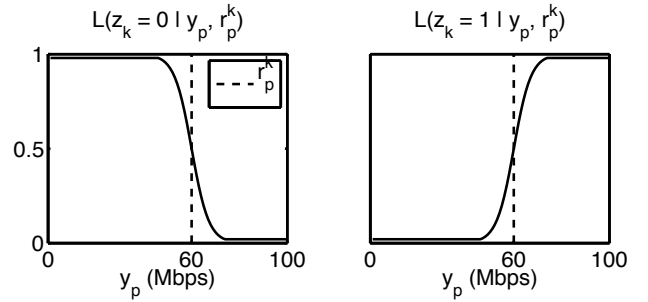}
\caption{Two possible values for $f_{y,z_k}$ representing the knowledge about the PAB of path $p$ obtained from a measurement $r_p^k = 60$Mbps.\label{fig:measurement_msg_imc}}
\end{figure}

As depicted in Fig.~\ref{fig:factor_graph}, after each measurement we add a function node $f_{z_k^p}$ to the factor graph and connect it with an edge to the variable node $y_p$ of the path that was probed.  There are two possible likelihood forms, depending on the outcome of the measurement; they are displayed in Fig.~\ref{fig:measurement_msg_imc}.  

\begin{figure}[!h]
\centering
\includegraphics[width=\linewidth]{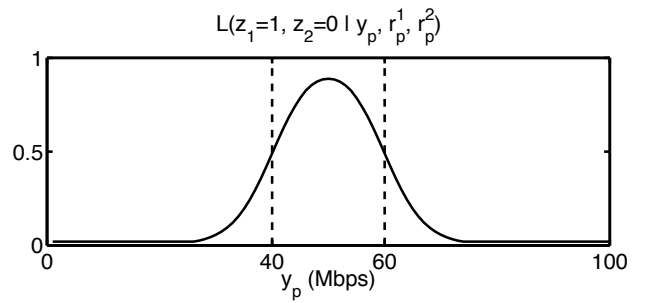}
\caption{Knowledge about a path $p$'s PAB from two measurements: 1) $r_p^1 = 40$, $z_1 = 1$, 2) $r_p^2 = 60$, $z_2 = 0$.\label{fig:two_measurements_imc}}
\end{figure}

If $z_k = 0$ then the probing rate is smaller than the PAB: $$f_{y_p,z_k} = {\mathrm{logsig}}(-\alpha(r_p^k-y_p)).$$  On the other hand, if $z_k = 1$ then $r_p > y_p$ and $$f_{y_p,z_k} = 1 - {\mathrm{logsig}}(-\alpha(r_p^k-y_p)).$$  The product of all the $f_z$ factor nodes for a path represents the cumulative knowledge obtained from measurements on this path.  In Fig.~\ref{fig:two_measurements_imc}, we show the product of two likelihood functions resulting from two measurements made at path $p$, one at $r_p^1 = 40$ and one at $r_p^2 =60$.  

\subsection{Active Sampling}

The estimation of available bandwidth based on self-induced congestion is an iterative process.  At every iteration, the probing rate is chosen according to some rules.  In the case of network-wide estimation, we must also determine which path to probe.  The possible sampling rules used to make these selections can be divided in two groups: adaptive (active) or non-adaptive (passive).  Non-adaptive sampling means that the sequence of measurements is pre-determined; the probing rate at step $k$ is not affected by previous measurements.  These strategies are simple and easy to implement, but can be inefficient.  Adaptive (active) selection algorithms, which use information extracted from previous measurements to make decisions about the future, can provide important reductions in the number of probes.  

\subsubsection{Path Selection}

We now describe two greedy active learning procedures to select the path to probe at each iteration (line 3 in Fig.~\ref{fig:alg_main}). Both algorithms are probabilistic in nature: they determine the probability that each path is chosen, and then the choice is accomplished by making a random selection according to the specified probabilities.

The first algorithm is called weighted entropy (WE).  For each path, we can calculate the entropy of the marginal posterior distribution of its PAB.  The entropy is an indication of the uncertainty associated with the current estimate; so WE assigns a probability that a path is selected is proportional to the entropy of the distribution.
The second algorithm, called weighted confidence interval (WCI), assigns a selection probability to each path that is proportional to the size of the current confidence interval $\beta_p$ of the path's PAB; it then chooses a path at random according to the assigned probabilities. 
In both algorithms, paths are more likely to be probed if there is more uncertainty about their PABs and the probability of probing a path that already satisfies our stopping criteria ($\beta_p \leq \beta$) is zero.



\subsubsection{Rate Selection}
	
To decide on the probing rate (line 4 in Fig.~\ref{fig:alg_main}), previous estimation tools either use deterministic binary search or simply increase the probing rate (linearly or exponentially) until it is greater than the available bandwidth.  
Our Bayesian framework allows us to adopt a more efficient and informative approach.   We choose the rate that bisects the marginal posterior distribution of the path.
By probing at the median, there is equal probability (according to our current knowledge) that the binary outcome will be $z_k = 1$ or $z_k=0$.  
We therefore maximize the expected information gain from our measurement; it is equivalent to conducting a probabilistic binary search for the available bandwidth on path $p$~\cite{cas:07}.  
By using a probabilistic rather than deterministic approach in rate selection, hard decisions (which could be incorrect) are not enforced.

\section{Results and Discussion}
\label{sec:res}

\subsection{Path Selection Simulations}
\label{ssec:sim}

The purpose of the simulations described in this subsection is to assess the efficacy of our proposed active sampling strategies. These are not network simulations, so they do not test modelling assumptions at all (that is the purpose of the simulations in Sec.~\ref{ssec:robust} and the online experiments in Sec.~\ref{ssec:exp}).  

We use the HOT topology generated using Orbis\footnote{\url{http://www.sysnet.ucsd.edu/~pmahadevan/topo_research/topo.html}}, which includes 939 nodes (896 end nodes) and 988 links.  From this set of links and nodes, we construct a distance matrix between all the nodes using shortest path routing and identify 2232 paths (source-destination pairs) that consist of at least seven links.  For our simulations, we wish to test our algorithm on topologies of different sizes and vary the number of paths over the range $M=50,100,150,200,250$.  For each value of $M$, we randomly select ten different subsets of $M$ paths from the entire set of 2232 paths.  For each of these 50 topologies, we assign link PABs using a uniform distribution between $[1,100]$ and repeat this process ten times to generate a total of 500 topologies.

At each iteration, probe outcomes are generated according to the likelihood model we constructed empirically in Sect~\ref{sssec:likelihood} ($\alpha = 0.28$) for $\epsilon=5$. For all simulations, $\gamma = 0.5$, which means that the value of the likelihood function at $y_p = r_p$ is $0.5$.  We compare three path selection algorithms (Round Robin (RR), WE and WCI) and also show the average number of measurements and accuracy required when our active learning algorithm is run independently and sequentially on each path (SEQ).  We use different values of $\beta$ and $\eta$ as stopping criteria; the algorithm stops when the size of the confidence interval $\beta_p$ is smaller than $\beta$ for all paths $p$.  If these conditions are not met, the algorithm stops after 10000 iterations. 

\begin{figure}[!t]
	\centering
	\includegraphics[width=\linewidth]{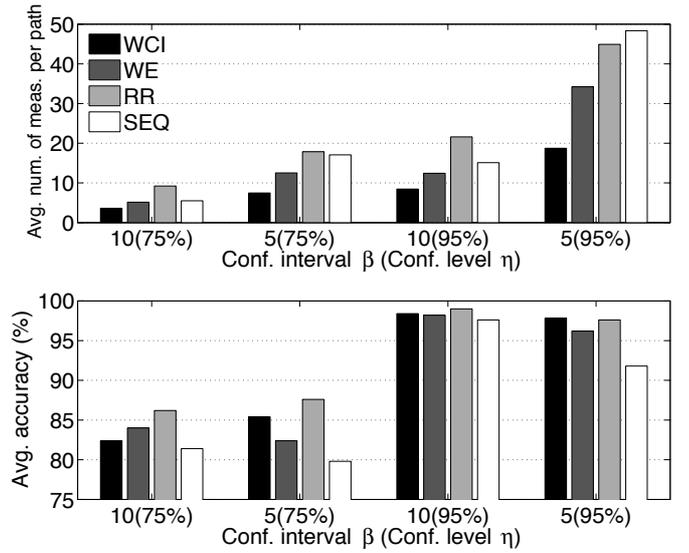}
	\caption{Simulation results: measurements required and accuracy achieved. Results are averaged over 500 topologies of various sizes for different confidence levels $\eta$ and intervals $\beta$.}
	\label{fig:simulation_sig}
\end{figure}

Fig.~\ref{fig:simulation_sig} shows the number of measurements per path required for the algorithm to terminate, as well as the accuracy (an estimate is considered accurate if the real PAB lies within the confidence limits: $\beta_{min} \leq y_{p} \leq \beta_{max}$).  In most cases, SEQ requires fewer measurements than the round-robin strategy with the graphical model.  This is due to the fact that not all paths require the same number of measurements.  In the RR case, the algorithm iterates through all paths, including those that have already met the required confidence criteria, which is not the case in SEQ.  Both data-driven approaches, WCI and WE, significantly reduce the number of measurements required while achieving satisfactory accuracy (i.e., the accuracy exceeds the requested confidence level $\eta$).

\begin{figure}[!h]
	\centering
	\includegraphics[width=0.9\linewidth]{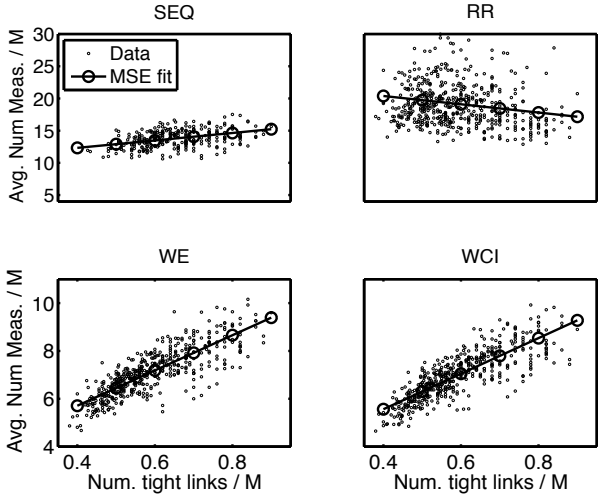}
	\caption{Simulated average number of measurements as a function of the number of tight links in the topology.  Both values are normalized by the number of paths $M$.  We show all the simulated values and a first degree polynomial fit for each technique.}
	\label{fig:simulation_info}
\end{figure}

We investigate the number of iterations for the case where $\eta = 0.95$, $\beta = 10$ in Fig.~\ref{fig:simulation_info}; we show the average number of measurements per path as a function of the number of tight links per path in the network.  
Due to the nature of our model, we can identify the PAB of each path if we know the PAB of all the tight links in the network.  
Therefore, we expect to make greater savings in terms of number of probes when the total number of tight links is small relative to the total number of paths (or, in other words, when the number of paths that share a single tight link is high).  
The average number of measurements per path required by WCI is between $46-73\%$ lower than the number required by RR and $39-55\%$ lower than SEQ.
WE and WCI provide important savings in terms of time and measurements without affecting the accuracy, but since WCI is slightly better in terms of average number of measurements, we use WCI for our online experiments.  As expected, when tight links are located on non-shared links, more measurements are required to achieve the same level of accuracy.

\subsection{Topology Accuracy Simulations}
\label{ssec:robust}

Our methodology assumes that the logical topology is known and stable during the estimation procedure.  
We are interested in assessing the robustness of our approach relative to i) errors introduced in the physical topology extraction using \texttt{traceroute} 
and ii) changes in the real topology in the middle of the estimation.

Let TE be the probability that path $p$ is incorrectly extracted using \texttt{traceroute}.
For each erroneously extracted path, there is a probability $q_{flip}$ that each link in the set $\mathcal{L}$ is mistakenly identified as either present or missing from path $p$\footnote{This probability is chosen such that the average path length remains constant.  Based on the topologies we used for our simulations, this probability depends on the number of links in the network and varies between 1-3\%.}.  More concretely, for each row of $\mathbf{P}$, there is a probability TE that every column entry is flipped with probability $q_{flip}$.  The result is a noisy factor graph (path matrix) that propagates inaccurate information because of invalid edges between path and link variable nodes.

\begin{figure} [!h]
\centering
\includegraphics[width=\linewidth]{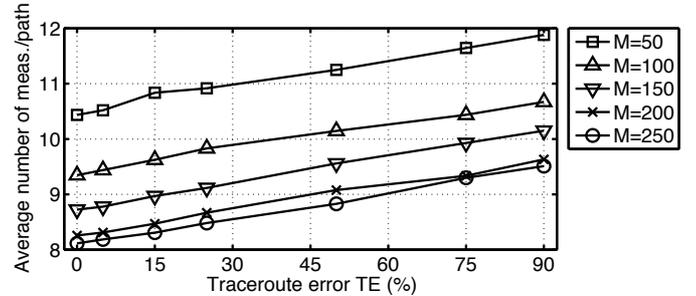}
\caption{Average number of measurements per path as a function of the traceroute error for topologies with different number of paths $M$.\label{fig:nm_large_v3}}
\end{figure}

For each of the 500 topologies we used in Sec.~\ref{ssec:sim}, we generate seven topologies by varying $TE$ over the range $0\%, 5\%,15\%,25\%,50\%,75\%, 90\%$.  
For the simulations, we use WCI for path selection, the same likelihood model with $\alpha=0.28$ and set $\gamma = 0.5$, $\epsilon = 5$ Mbps, $\eta = 0.95$, $\beta = 10$ Mbps, $B_{min} = 1$ Mbps, $B_{max} = 100$ Mbps. 
In Fig.~\ref{fig:nm_large_v3}, we show the average number of measurements per path as a function of TE. 
As expected, the number of iterations required to achieve the requested confidence level and tightness increases for topologies with a greater probability of \texttt{traceroute} error.
However, this augmentation is not significant; even with $TE=90\%$, the estimation requires only 1.5 more measurements per path on average.

\begin{figure} [!h]
\centering
\includegraphics[width=\linewidth]{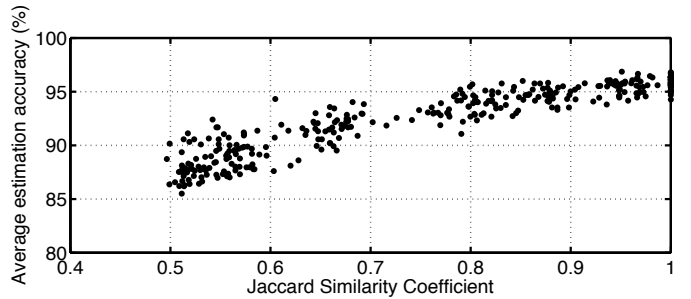}
\caption{Average estimation accuracy (Jaccard Similarity Coefficient = $|A \cap B| / |A \cup B|$) as a function of the topology accuracy for all topologies. \label{fig:acc_large_v3}}
\end{figure}

To quantify the similarity between topologies and provide a more meaningful metric than TE, we use the Jaccard similarity coefficient. It is equal to the size of the intersection (number of correctly identified links) divided by the size of the union (all links from both topologies)~\cite{jac:1901}.
We display the average accuracy of our estimates over all topologies in Fig.~\ref{fig:acc_large_v3}.  
Our simulation results show that, for topologies of any size, as long as the \texttt{traceroute} methodology produces a matrix $\mathbf{P}$ with a similarity coefficient greater than $0.5$, $85\%$ of the paths are estimated accurately on average.  Therefore, even when it uses an inaccurate path matrix, our methodology can generate reasonably precise estimates without any significant inflation in the number of probes required.

As far as topology stability is concerned, we have not performed any simulations, but we have studied empirically the validity of our assumption.  
Before each of our online experiments, we generated the matrix $\mathbf{P}$ and studied its similarity with previous matrices for the same set of nodes.
We conclude that the PlanetLab network is stable enough to assume that topologies remain constant during the estimation procedure (\citet{son:07} made similar observations).  Although it is probably safe to assume that the topologies are constant for even a longer period of time (at least 24 hours from our observations), we continue to generate a new matrix $\mathbf{P}$ before every experiment since it is neither time nor resource consuming.  It is important to note that the logical topology is not always affected by variations in the physical topology.  Therefore, they do not necessarily imply modifications in the path matrix and the associated factor graph.  

\subsection{Online experiments}
\label{ssec:exp}

For our online experiments, we have deployed our measurement software coded in C on various nodes on the PlanetLab network\footnote{Although the PlanetLab (\url{http://www.planet-lab.org/}) network was once believed to be too heavily loaded, \citet{spr:06} explained that PlanetLab has evolved and this is no longer true.}.  
We use a topology with six nodes\footnote{planetlab3.csail.mit.edu, planetlab-1.cs.unibas.ch, planlab1.cs.caltech.edu, planetlab2.acis.ufl.edu, planetlab1.cs.stevens-tech.edu, planetlab2.csg.uzh.ch.}, $M=30$ paths and $N=65$ logical links.  For all our experiments, the likelihood model is the one presented in Sec.~\ref{sssec:likelihood} (with $\epsilon=5$ and $\alpha=0.28$) and WCI is used to select the path to probe at each iteration.  Also, we choose $B_{min} = 1$ Mbps and $B_{max} = 100$ Mbps as conservative estimates of the PAB of each link (we assume that the links with the highest capacity are 100 Mbps links).

Each run includes an estimation of all the paths followed by a testing procedure.
The estimation terminates when the stopping criteria, $\beta = 10$ Mbps and $\eta = 0.95$, are met for all paths.
We validate our results by sending trains of 2400 packets of 1000 bytes (the equivalent of 60 seconds of video encoded at 320 kbps) and observing the output rate.
For each run, we perform a total of 16 tests; four tests on four disjoint paths.  In each of the tests, the sending rate of the train is different --- the lower bound of the confidence interval $\beta_{min}$, the lower bound plus $\epsilon=5$ Mbps, the upper bound of the confidence interval $\beta_{max}$, and $\epsilon=5$ Mbps above the upper bound.  For each test, we compute the empirical probability that the output rate is within $\epsilon=5$ Mbps of the input rate ($z=1$).  

In this first experiment, we set $\gamma=0.5$ and wish to verify if the confidence intervals produced include the value of the PAB.  To do so, we compute the average over 20 runs of the empirical probability $\widehat{\Pr} (r'_p \geq r_p - \epsilon)$ for each one of the four tests.  
For the probes, we use $N_t=3$ trains per measurement, a packet size of $P_{size}=1000$ bytes and vary the number of packets in each train in the range $L_s=[25,50,100,150,200,250]$.  

\begin{figure}[!h]
	\centering
	\includegraphics[width=\linewidth]{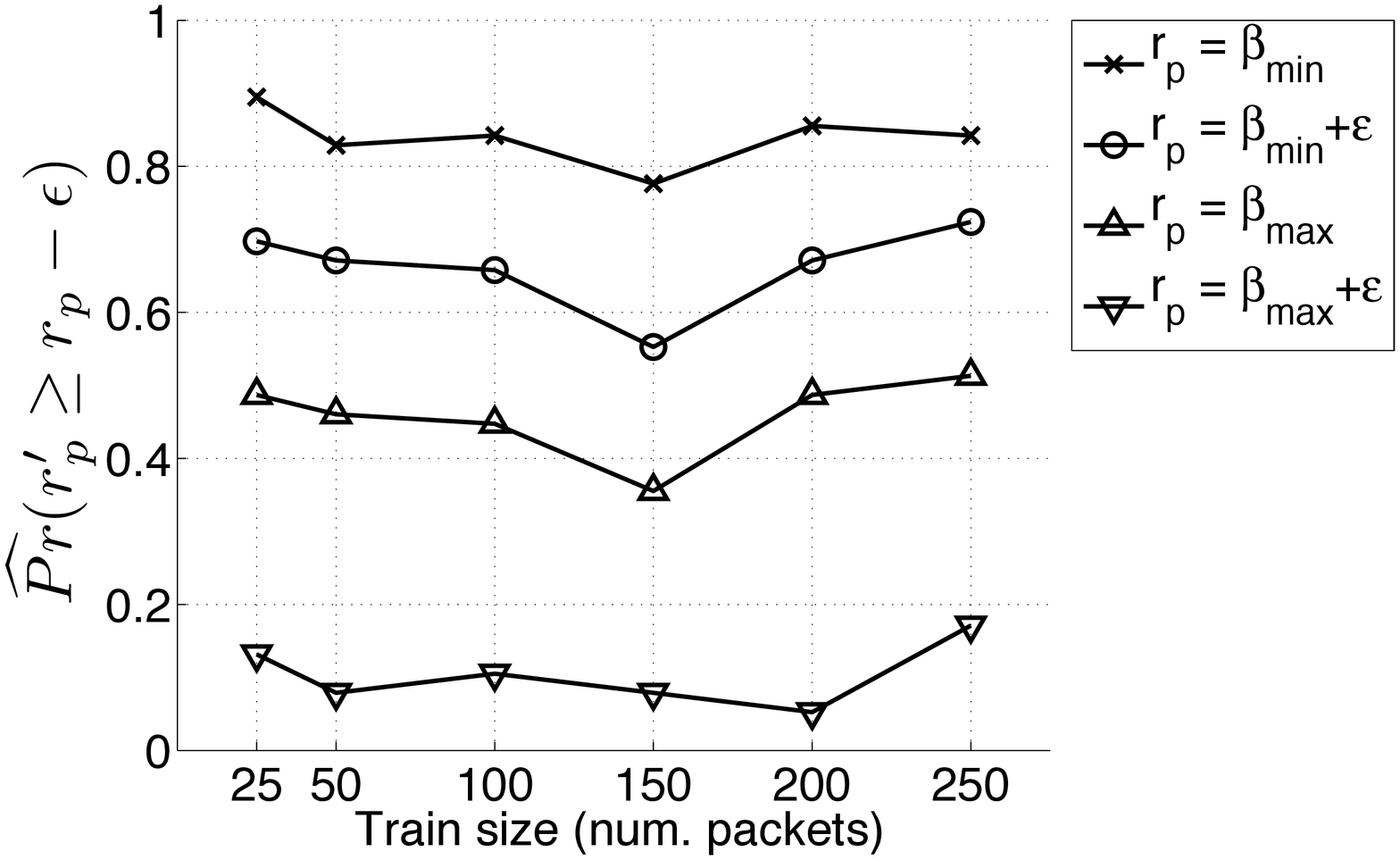}
	\caption{Empirical probability that the output rate is within $\epsilon$ Mbps of the input rate.  Each point represents the average of 80 test results (20 runs).\label{fig:live_accuracy_train_size_info}}
\end{figure}

In Fig.~\ref{fig:live_accuracy_train_size_info} we show the empirical probability (averaged over 80 tests) that the output rate is within $\epsilon=5$ Mbps of the input rate for four different probing rates relative to the confidence interval bounds.  The first observation is that the number of packets used in trains induces very little variation in empirical probability for all the probing rates.  This suggests that, for this network at least, 25 packets per train would suffice.  For all the train sizes we tested, the desired probability $\gamma=0.5$ is included in the probability interval of $\beta_{min}$ and $\beta_{max}$.  This result confirms that our method is able to produce intervals that include the value of the PAB accurately.  The fact that $\gamma=0.5$ is very close to the upper bound suggests that we might underestimate the PAB.  We discuss possible reasons for this below.

\begin{figure}[!h]
	\centering
	\includegraphics[width=\linewidth]{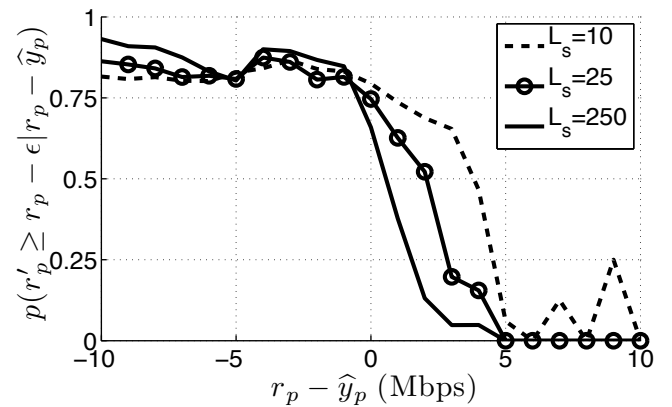}
	\caption{Empirical probability of observing $z=1$ averaged over 17966 measurements as a function of the difference between the probing rate and the estimated PAB (MAP of the marginal posterior).\label{fig:raw_trainsize_info}}
\end{figure}	

We investigate the impact of the train size by using the raw data collected at each node during the 20 runs (18000 measurements for each value of $L_s$).  In Fig.~\ref{fig:raw_trainsize_info}, we show the average empirical probability of observing $z=1$ as a function of the difference between the probing rate and our estimate of the PAB (here we use the marginal maximum a posteriori (MAP) estimate).  Since we set $\gamma=0.5$, we expect the probability of observing $z=1$ to be near $0.5$ when the probing rate is equal to the PAB ($r_p - \hat{y}_p = 0$).  However, what we observe is that the probability is closer to $0.75$ at that point, which is approximately the average empirical probability at $\beta_{min} + \epsilon$ in Fig.~\ref{fig:live_accuracy_train_size_info}.  This confirms a slight underestimation of the PAB, which is probably due to an inaccurate likelihood model.  The figure also shows that as the train size is reduced, the measurements become more noisy and the bias (underestimation) becomes more significant.  

\begin{figure}[!h]
	\centering
	\includegraphics[width=\linewidth]{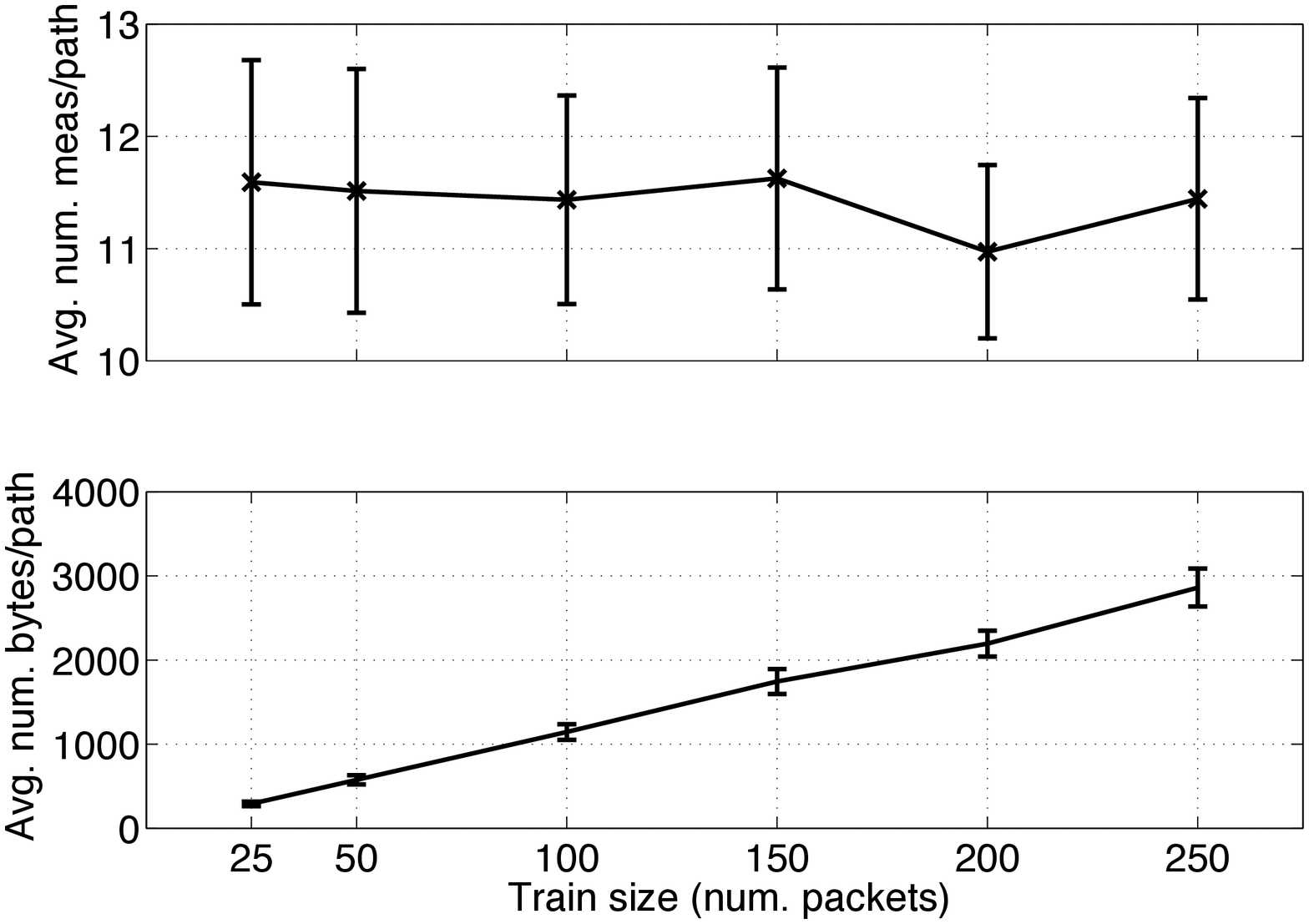}
	\caption{Number of measurements (TOP) and bytes (BOTTOM) used per path (averaged over 20 runs for each train sizes $L_s$) during the estimation procedure.  \label{fig:bytes_per_trainsize_info}}
\end{figure}

Figs.~\ref{fig:live_accuracy_train_size_info} and~\ref{fig:raw_trainsize_info} indicate that the accuracy obtained when using $L_s=25$ and $L_s=250$ packets is similar.  
In Fig.~\ref{fig:bytes_per_trainsize_info}, we show the average number of measurements and bytes per path required to complete the estimation procedure as a function of the train size.  Since the number of measurements is constant for all values of $L_s$, we observe a linear growth in the number of bytes required to achieve the desired accuracy.  From these results, it is now clear that using 25 packets per train is optimal as it provides similar accuracy to larger train sizes with significant savings in terms of number of probes.


In the previous experiment, where $\gamma=0.5$, the probability of observing $z=1$ when the input rate is equal to the lower bound of our confidence interval is $0.875$.  That probability drops to $0.7$ for the rate at the middle of our confidence interval ($\beta_{min} + \epsilon$).  We perform another experiment of 20 runs with $L_s=25$ and $\gamma=0.9$.  By increasing $\gamma$, we obtain higher guarantees for rates at the lower bound ($0.97$) and in the middle ($0.86$) of the confidence interval.
However, increasing the value of $\gamma$ results in a larger number of measurements.  
For $\gamma=0.9$, the average number of measurements per path was $33\pm1$ (compared to $12\pm1$ for $\gamma=0.5$) an augmentation of $175\%$).  

\begin{figure}[!h]
	\centering
\includegraphics[width=\linewidth]{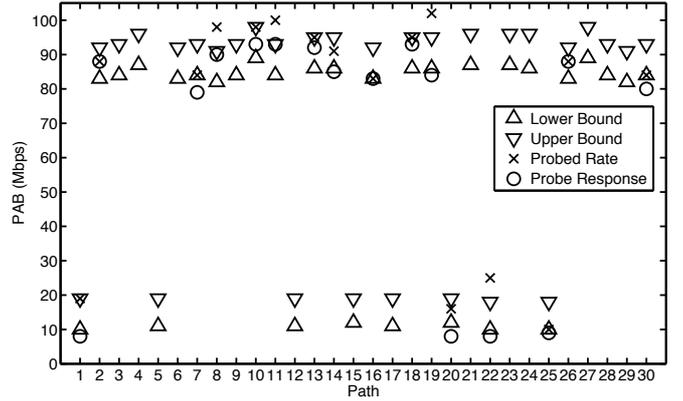}
	\caption{Bounds of the confidence intervals for a 30 paths topology in a sample run performed for $L_s=25$ and $\gamma=0.5$.	\label{fig:pl_exp_info}}
\end{figure}

In Fig.~\ref{fig:pl_exp_info}, we display the confidence intervals as well as the test results (probe rate and output rate) for one of the runs performed with $L_s=25$ and $\gamma=0.5$.  The outcome of this particular run demonstrate the clear heterogeneity of the PlanetLab network; over $25\%$ of the paths have small (less than $20$ Mbps) PAB whereas the other $75\%$ have PAB greater than $80$ Mbps.  The tight links on the paths with lower PAB could either be heavily utilized $100$ Mbps links or, more likely, $10$ Mbps links with small amounts of cross-traffic.  These findings about the PlanetLab network correspond to those of~\citet{lee:05}.

%

\begin{table}[!h]
	\caption{Average time and bytes used by Pathload and our approach for $M=30$ paths topology over 5 runs.\label{tab:pathload_info}}
	\centering
	\begin{tabular}{c|cc}
		& seconds / path & kbytes / path \\
		\hline
		Pathload & $27.0 \pm 0.8$ & $10806 \pm 1058$\\
		Our Approach & $7.1 \pm 0.3$ & $612 \pm 18$
	\end{tabular}
\end{table}

It is interesting to compare our estimation methodology to another tool based on the classical definition of available bandwidth to examine the extent of correlation between the two metrics.  We choose to compare our results with those obtained using Pathload (version 1.3.2)~\cite{jai:03} because it is known to be very accurate.  Using the same topology described above ($M=30$, $N=65$), we run both Pathload sequentially on every single path and our algorithm (WCI and $L_s=25$, $\gamma=0.5$).  Since the metrics are different, a complete correspondence between the estimates is not expected. Nonetheless, both estimation techniques strive to examine the same path property (at what rate can probe trains be sent without inducing congestion).  The confidence intervals obtained from both tools overlap for $53\%$ of the paths -- $76\%$ if we tolerate a 2Mbps error.  This highlights the fact that there is a certain level of correspondence between the two metrics.  Table~\ref{tab:pathload_info} compares  the number of bytes transmitted and time elapsed.  We can see that our approach provides significant gains in terms of measurement latency ($75\%$ savings) and overhead ($95\%$ savings).  

Comparing the overhead of our technique with Pathload's confirms that previous tools are not well suited to multi-path estimation.  The only other approaches that can produce efficient network-wide AB estimates are BRoute~\cite{hu:05} and bandwidth landmarking~\cite{man:07bandwidth}.  In both cases, there is very little details on the actual overhead incurred by their techniques.  \citet{hu:05} claim that $80\%$ of the available bandwidth estimates obtained from BRoute are accurate within $50\%$ when using a subset that includes only $7\%$ of all paths.  However, there is no mention of how many measurements are required for each path.

\section{Conclusion}
\label{sec:conclusion}

In this paper, we presented a novel technique based on a probabilistic framework to estimate network-wide probabilistic available bandwidth.  We introduced PAB, a new metric with adjustable parameters that addresses issues related to the dynamics and variability of available bandwidth.  Our methodology based on factor graphs and active sampling is the first to combine both techniques in the context of available bandwidth estimation.  To further reduce the overhead of our technique, we are currently working on a new measurement strategy and likelihood model based on chirps rather than trains of packets, which, from our preliminary results, can achieve significant savings in terms of probing overhead.

\bibliographystyle{elsarticle-num-names}
\bibliography{availbw}

\end{document}